%% file: ms.tex
\documentclass[12pt,preprint]{aastex}

\def\vlsr{$V_{\mbox{\scriptsize LSR}}$}
\def\kms{~km~s$^{-1}$}

\def\etal{~et\ al.\ }
\def\h2o{H$_{2}$O}
\def\deg{$^{\circ}$\hspace{-2pt}}
\def\w3{W3~IRS~5}

\shortauthors{H.~Imai \etal}
\shorttitle{Linear Polarization of Water Masers in \w3}

\begin{document}

\title{Linear Polarization Observations of Water Masers in \w3}

\author{Hiroshi Imai\altaffilmark{1,2,3}, Shinji Horiuchi\altaffilmark{4,5}, 
Shuji Deguchi\altaffilmark{6} and Osamu Kameya\altaffilmark{1,2}}

\altaffiltext{1}{Mizusawa Astrogeodynamics Observatory, National Astronomical 
Observatory, Mizusawa, Iwate 023-0861, Japan}
\altaffiltext{2}{VLBI Exploration of Radio Astrometry Project Office, National 
Astronomical Observatory, Mitaka, Tokyo 181-8588, Japan}
\altaffiltext{3}{Joint Institute for VLBI in Europe, Postbus 2, 
7990 AA Dwingeloo, the Netherlands}
\altaffiltext{4}{VLBI Space Observatory Programme Project Office, 
Natinal Astronomical Observatory, Mitaka, Tokyo 181-8588, Japan }
\altaffiltext{5}{Jet Propulsion Laboratory, 4800 Oak Grove Drive, 
Pasadena, CA 91109, USA}
\altaffiltext{6}{Nobeyama Radio Observatory, National Astronomical Observatory, 
Minamimaki, Minamisaku, Nagano 384-1305, Japan}

%E-mail address:
%H. Imai; imai@jive.nl, S. Horiuchi; shoriuchi@mail1.jpl.nasa.gov,	
%S. Deguchi; deguchi@nro.nao.ac.jp
%O. Kameya; kameya@miz.nao.ac.jp

%\received{?? April 2003}
%\accepted{??? ????? 2003}
%\journalid{???}{?? ???? 2003}
%\articleid{??}{??}

\begin{abstract}

We present a magnetic field mapping of water maser clouds in the 
star-forming region \w3, which has been made on the basis of the linear 
polarization VLBI observation. Using the Very Long Baseline Array (VLBA) 
at 22.2~GHz, 16 of 61 detected water masers were found to be linearly 
polarized with polarization degrees up to 13~\%. Although 10 
polarized features were widely distributed in the whole \w3 water maser 
region, they had similar position angles of the magnetic field vectors 
($\sim$ 75\deg\ from the north). The magnetic field vectors are roughly 
perpendicular to the spatial alignments of the maser features. They are 
consistent with the hourglass model of the magnetic field, which was 
previously proposed to explain the magnetic field in the whole W3~Main 
region ($r\sim$~0.1~pc). They are, on the other hand, not aligned to the 
directions of maser feature proper motions observed previously. This 
implies that the \w3 magnetic field was controlled by a collapse of the 
W3~Main molecular cloud rather than the outflow originated from \w3. 
\end{abstract}


%\newpage
\section{Introduction}

Birth of massive stars is characterized by a collapse of molecular cloud 
core when self-gravity overwhelms the magnetic pressure in the core 
as a result of magnetic field dilution. In this situation, the core is not 
supported by the magnetic field pressure in the cloud core any more, and 
called to be magnetically super-critical (e.g., \citealt{shu87,fei99}, and 
references therein). Measurements of the magnetic field strengths and 
directions in star-forming regions have been made in a variety of spatial 
scales, from whole star-forming regions (1--10~pc) down to individual 
molecular cloud cores (e.g., \citealt{hei93,mck93,eva99}, and references 
therein). In the smaller scale, the measurements were made mainly with 
radio interferometers with higher angular resolution. However, because 
of complicated structures of the magnetic fields expected from 
the presence of the dense clumps and outflows, and because of the 
uncertainty in calibration in the polarimetric interferometry, it is more 
difficult to map the magnetic field accurately in this scale. 

Water maser emission is often associated with star-forming regions (SFRs) 
with energetic outflows from young stellar objects (YSOs) (e.g., \citealt
{eli92a}a, b and references therein). VLBI observations have revealed that 
water maser emission consists of clusters of compact maser features with 
a typical size of 1~AU and a velocity width of 1\kms (e.g., \citealt{rei81}). 
Measurements of Doppler velocities and proper motions of maser features 
have led to the conclusion that the spatial motions of maser features really 
reflect the true three-dimensional gas kinematics (e.g., 
\citealt{gwi92,tor01a}a, b). Previous observations have found expansion 
motions produced by outflows from YSOs or expanding H{\rm II} regions 
(e.g., \citealt{ima00}, hereafter Paper {\rm I}, and references therein). 
Shocks, which have velocities exceeding about 20\kms\ and are running 
into high-density 
($n_{\mbox{\tiny H$_{2}$}}\: \geq$~3$\times$10$^{6}$ cm$^{-3}$) magnetized 
material, successfully explain the water maser emission associated with 
energetic YSO outflows (e.g., \citealt{hol89, eli89}, 1992; \citealt{hol93}). 

Polarimetric mapping of individual water maser features were made for the 
first time using VLBI data in W51~M by \citet{lep98} (hereafter LLD). They 
found that the observed directions of linear polarization were well aligned 
along a stream of water maser features found in W51~M. It should be noted 
that the observed direction of the linear polarization does not straightly 
indicate the magnetic field direction of the cloud because it depends on 
the angle between the magnetic field and the photon propagation direction 
(toward an observer) (e.g., \citealt{gol73,deg86b,deg90}). In the case of 
W51~M, the observed directions of linear polarization were interpreted 
to be parallel to the magnetic field vector projected on the sky plane because 
the field vector is expected to be roughly tangential to the line of sight due 
to coupling with the Sagittarius spiral arm. LLD concluded that the alignment 
of the water maser linear polarization was created by shocks caused by the 
nearby expanding H{\rm II} region. The linear polarization of water masers 
as a tracer of the interstellar magnetic fields is indirectly supported by that 
the linear polarization directions in a maser region are stable on a timescale of 
several years despite of much shorter time scale of lifetimes and flux 
variability of maser features (\citealt{abr94}; LLD; \citealt{hor00a, hor00b}; 
Horiuchi \& Kemaya 2003 in preparation).

\begin{figure*}
\epsscale{0.4}
\plotone{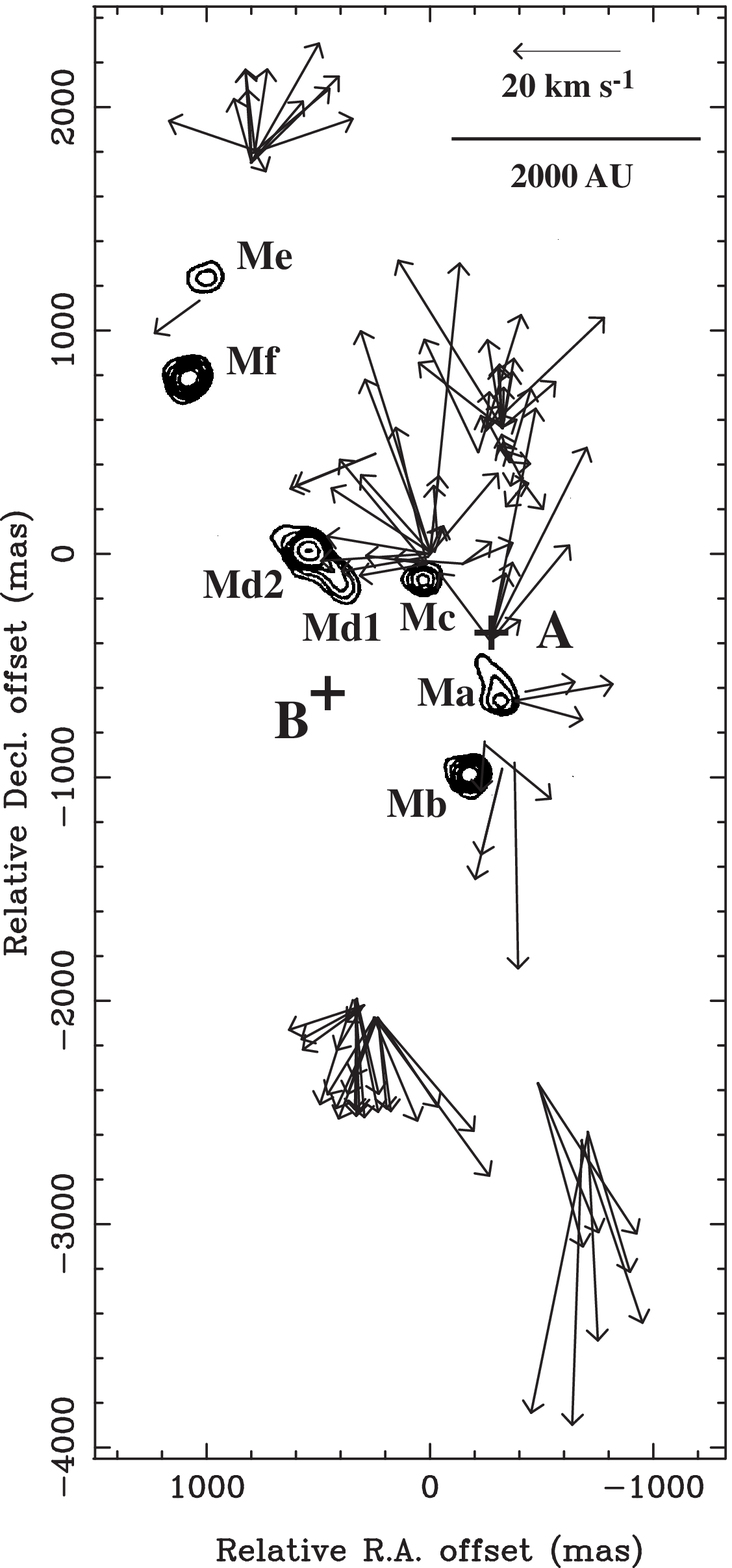}

\caption{Water maser features and compact radio continuum sources in 
\w3. Arrows indicate locations and proper motions of maser features, 
which were revealed by Paper {\rm I}. The arrow length indicates the 
magnitude of the proper motion per year ($\times $325), respectively. 
The radio continuum sources, which were mapped by Claussen\etal(1994), 
are shown as contours. The designations the continuum sources are 
same as those in Tieftrunk\etal(1997). Two plus marks A and B indicate 
the locations of driving sources of the outflows in \w3, which were 
estimated by Paper {\rm I}. 
\label{fig1}}

\end{figure*}

This paper discusses the results of magnetic field mapping of water maser 
clouds in the massive-star forming region \w3 revealed by a VLBI observation. 
\w3 is considered to be at the very early phase of star formation because of 
lack of high velocity components in the water masers, as suggested by 
\citet{gen77}. Previous observations also supported this interpretation on 
\w3 (e.g., \citealt{cla94, tie97, rob97}). In a 2\arcsec$\times$3\arcsec\ 
field of \w3, \citet{cla94} and \citet{tie97} found the distribution of a 
cluster of centimeter-continuum emission sources. 
\citet{ima00} investigated the 3-D kinematics of the water masers in 
\w3\ and estimated a distance to this region to be 1.83$\pm$0.14~kpc. 
In the present paper, the value of 1.8~kpc is adopted as the distance to 
\w3. \citet{ima02} (hereafter Paper {\rm II}) investigated the physical 
condition of the region where the water masers are excited. They found 
that turbulent motions dominate the morphology and kinematics on 
a microscopic scale (down to 0.01~AU). Figure \ref{fig1} 
shows the distribution and proper motions of the water maser features 
found in Paper {\rm I}. 

\citet{tro89} and \citet{rob93} (hereafter RCTG) measured the Zeeman effect 
of the H{\rm I} line in \w3. They revealed the magnetic field projected in the 
line of sight on a scale of $\sim$~0.6 pc, and demonstrated an "hourglass" 
model of the magnetic field. The above scale is roughly equivalent to the core 
size found in this cloud found in the $^{13}$CO line by \citet{hay89}. 
A pinch of the hourglass was found to be located very close to \w3 region 
and to have the magnetic field strength higher than 1~mG. \citet{gre94} also 
made a polarimetric observation at sub-millimeter wavelengths and estimated 
the magnetic field in \w3\ on a scale of 0.1~pc; the result supports the 
hourglass model. \citet{bar89} found that the \w3 water masers were highly 
linearly polarized. \citet{sar01} (hereafter STR) and \citet{sar02} (hereafter 
STCR) measured the Zeeman effect of the bright components of the water 
masers and estimated the magnetic field strength along the line of sight 
to be $\sim$~30~mG.

In the present paper, we focus on an issue whether or not the magnetic 
field measured in water masers traces that found on the larger scale. The 
cause that determines the shape of the large-scale magnetic field is also 
speculated: either the collapse of the parent molecular cloud or the formation 
of a magneto-hydrodynamical jet. Section 2 describes the polarimetic 
observations using the Very Long Baseline Array (VLBA) and the Very Large 
Array (VLA) of the National Radio Astronomy Observatory (NRAO)\footnote[1]
{The NRAO is a facility of the National Science Foundation of USA, operated 
under a cooperative agreement by Associated Universities, Inc.} and data 
reduction. The results are presented in \S  3. Section 4 discusses mainly the 
relation between the measured directions of maser linear polarization and 
possible directions and strength of the magnetic field in the observed scale 
($<$5\arcsec) of \w3.

\placefigure{fig1}

\section{Observations and data reduction}
\label{sec:observation}

The polarimetric observation of water masers in \w3 was made on 1998 
November 21 using 10 telescopes of the VLBA. The water masers and 
calibrators (NRAO~150, JVAS~0212$+$735, and 0423$-$014) were observed 
for 10~hrs in total. The observed signals were recorded in a 4-MHz base band 
with both right and left circular polarization (RCP, LCP) in 2-bit sampling. 
The recorded data were correlated with the VLBA correlator in Socorro, 
which yielded parallel and cross-hand visibilities with 128 spectral channels. 
The correlated data covered a velocity range of 
$-$58\kms\ $\leq$ \vlsr $\leq$ $-$5\kms\ and a velocity spacing of 
0.42\kms\ per spectral channel. 

Data calibration and imaging were made using the procedures in 
AIPS\footnote[2]{The NRAO 
Astronomical Imaging Processing System.} mostly. First, the data of the 
parallel-hand-polarization visibilities were reduced in the usual procedures 
(e.g., \citealt{rei95, dia95}) and fringe-fitting, self-calibration, and complex 
bandpass solutions were obtained independently in RCP and LCP. 
Although the water maser emission was strong enough to apply the "template 
method" to calibrate the visibility amplitudes \citep{dia95}, the band width 
of 4~MHz was too narrow to obtain a sufficient number of emission-free 
velocity channels in the template total-power spectrum of the maser emission. 
Instead, we applied tables of the measured system noise temperatures and 
antenna gains. In the self-calibration, we obtained a Stokes {\it I}\  maser map 
of the velocity component (maser spot) at $-$40.2\kms\ and complex gain 
solutions in every 6~s. 

Second, the instrumental polarization parameters (D-terms) of the telescopes 
were obtained using the AIPS RUN script CROSSPOL and the task LPCAL with 
scans of 0420$-$014, which is well linearly polarized. The D-term solutions were 
applied to the maser data. After the D-term calibration, the Stokes {\it I, Q, U}\  
maps of maser spots were obtained independently. The naturally-weighted 
synthesized beam was 0.68$\times$0.49 mas, with the major axis at a position 
angle (PA) $PA\:=$ 3\deg.3. No smoothing in velocity channel was applied. 
The detection limit of about 30 mJy~beam$^{-1}$ at a 5-$\sigma $ noise was 
obtained in velocity channels without bright maser emission. Before making 
the final image cubes, we looked for the masers by making wider maps covering 
a 2\arcsec$\times$5\arcsec\ field with a larger synthesized beam. 

Third, each velocity component (maser spot, see e.g., \citealt{gwi94a}a, b; 
Paper {\rm II}) in the image cube was fitted by a two-dimensional Gaussian 
brightness distribution using the AIPS task JMFIT. The uncertainty of the 
estimated position due to thermal noise was 3 microarcseconds ($\mu$as) 
for a 1-Jy spot. On the other hand, positional accuracy of the brightness peak 
of a maser feature (a cluster of maser spots) depends on the distribution of 
maser spots in the feature and the brightness distributions of the maser 
spots and limited to be typically 10 $\mu$as. The electric vector position 
angle (EVPA, see LLD), EVPA~$=$~arctan($Q/U$)/2 of each maser spot 
was calculated from the Stokes $Q$ and $U$ parameters. For each 
maser feature (a cluster of maser spots existing together within $~$~1~mas 
and $~$~1\kms\ in position and velocity, Paper {\rm I} and {\rm II}), the 
degree of linear polarization and the EVPA were obtained from the 
Stokes parameters ({\it I, Q, U}) of the intensity peak in the maser feature. 

The water masers were also observed with the VLA (26 antennas) in 
C-configuration on 1998 November 28, one week later the VLBA observation. 
The water masers and calibrators (3C~48, 0420$-$014, and 3C~138) were 
scanned for 2~hrs in total. The observed signals were recorded in a 3-MHz 
base band both in right and left circular polarization. The correlated data had 
64 spectral channels with a velocity spacing of 0.63\kms\ per spectral channel. 
The D-terms of the antennas were obtained from the scans of 3C~48, which 
covered a variation of parallactic angle over 180\deg, assuming the intrinsic 
EVPA of $-$72\deg.5. Then we confirmed that the obtained EVPAs of 3C~138 
and 0420$-$024 were consistent with those previously obtained by 
\citet{aku93} and \citet{mar02}, respectively. 

The EVPA at \vlsr$=$ $-$42.2\kms\ in the maser feature \w3:I2003 {\it 5} 
(see table \ref{tab1}) was compared between the VLBA and VLBA data. 
At this velocity, the maser spot was most prominent and its polarimetric 
parameters were well determined. A correction 
($\Delta$EVPA~$=$~39.\deg3) was applied to the EVPAs obtained with 
the VLBA. Taking into account the variation of EVPA in the maser 
feature ($\leq$ 3\deg), the ambiguity of the assumed EVPA of 
3C~48 ($\sim$~5\deg), and the accuracy of the calibration of the VLA 
data (uncertainty in R$-$L phase determination, $\leq$~0.3\deg), the 
accuracy of EVPA was expected to be $\sim$6\deg. Because the VLA 
observation was made 7 days after the VLBA observation, the estimated 
EVPAs might suffer from the EVPA time variation of the feature 
\w3:I2003 {\it 5}. 

\placetable{tab1}

\begin{figure*}
\epsscale{0.5}
\plotone{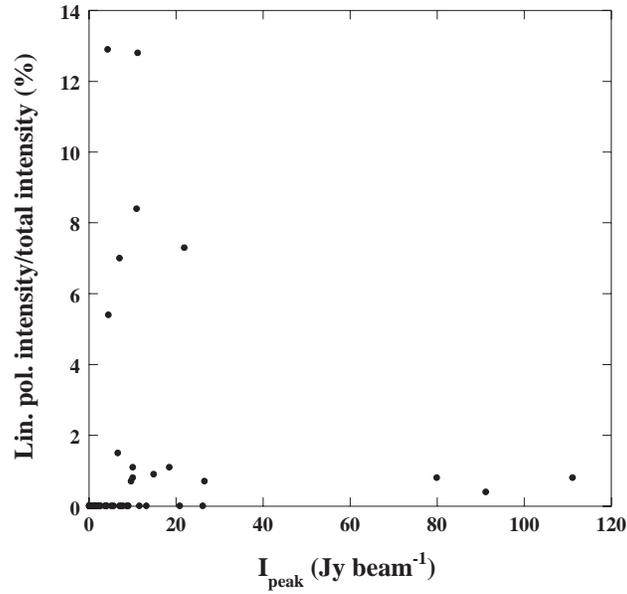}

\caption{Distribution of polarization degrees of maser features against 
their peak intensities. 
\label{fig2}}
\end{figure*}

\placefigure{fig2}

\placefigure{fig3}

\section{Results}

We detected 61 maser features consisting of 290 maser spots. These numbers are 
smaller by factors of 2--3 and 4--7, respectively, than those in the previous VLBA 
observations (Paper {\rm I}, see also Paper {\rm II}) while the sensitivity of the 
present observation was better than the previous ones. The decrease of features 
was likely due to less activity in maser excitation, while the decrease of spots was 
due to less velocity resolution by a factor of 4 in the present observation. 

\begin{figure}
\epsscale{0.5}
\plotone{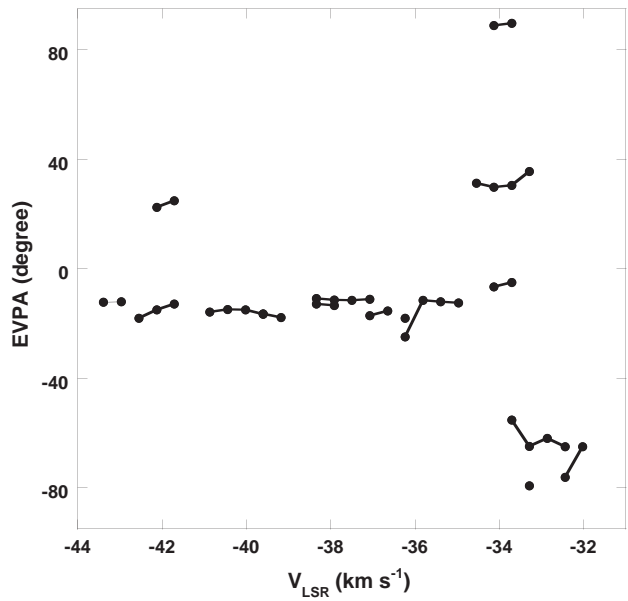}

\caption{Distribution of EVPAs of maser spots against their Doppler velocities.
A filled circle shows the maser spot detected its linear polarization. Maser spots 
in the same maser feature are connected each other with a thick line. 
\label{fig3}}
\end{figure}

Table \ref{tab1} summarizes the parameters of water maser features in \w3. 
Only for 15 of 61 features, the proper motions were measured previously 
in Paper {\rm I}. It was difficult to identify the same features between the 
observations of Paper {\rm I} and the present work because many features 
were likely to appear and disappear in the crowded regions during the epoch 
separation of 11 month. The highest degrees of polarization was 13\%  in the 
present observation, which was smaller than those (up to 30\%) found 
in a previous single dish observation \citep{bar89} and an unpublished VLBI 
observation with a 400-km baseline. As shown in figure 
\ref{fig2}, no correlation was found between a feature intensity and a degree 
of linear polarization. LLD found a systematic change of EVPAs in a single 
maser feature with the full velocity width $\Delta V\:=$~1.3\kms. On the other 
hand, as shown in figure \ref{fig3}, EVPAs were almost the same (within 
15\deg) in all of the features with measured linear polarization in \w3. 

\begin{figure}
\epsscale{0.9}
\plotone{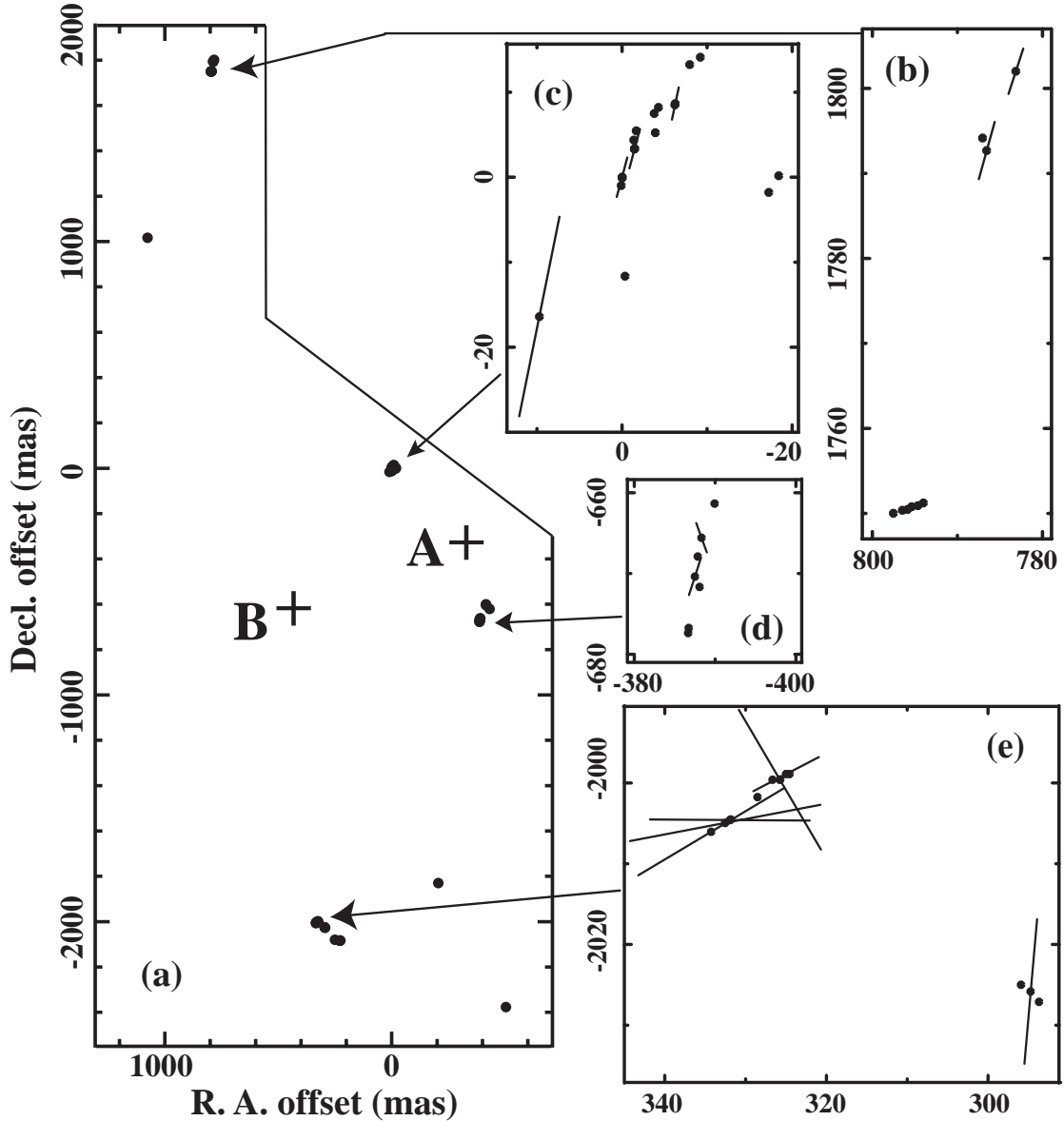}

\caption{Map of linearly-polarized maser features in \w3. (a): Locations of 
the maser features (filled circle) detected on 1998 November 21. Two 
plus marks are the same as those in figure 1. The locations of the maser 
features were estimated with respect to the driving sources of the outflows 
by assuming the stable distribution of clusters of maser features among 
observations made in the previous (Paper {\rm I}) and the present works. 
The location uncertainty of the A outflow will be within 100~mas. (b)-(e): 
Details of the feature distribution and the directions of linear polarization 
of the features. A length and a position angle of a bar indicate a logarithmic 
scale of the degree of linear polarization and an EVPA, respectively. 
\label{fig4}}
\end{figure}

\placefigure{fig4}

Figure \ref{fig4} presents the spatial distribution and the linear polarization 
directions of the maser features. The present maser map was easily overlaid 
on the previous ones in Paper {\rm I}. We found that a cluster 
of maser features is elongated in the north--west and south--east direction 
and that the radial velocities have been almost the same among previous 
maser maps of Paper {\rm I} and {\rm II} and the present one (figure 
\ref{fig4}c). Very likely this cluster corresponds to the cluster of maser 
features with the measured Zeeman effect (the features in figure 1 of STR, 
see also STCR). As STR mentioned, the cluster was located very close to the 
radio continuum source "c" (\citealt{cla94}, see in figure \ref{fig1}). It was 
quite difficult, however, to exactly identify the same features from one 
observing epoch to another because of too large time spans between two 
observations ($\gtrsim$~14 months).

The observed EVPAs were roughly the same among distant maser features;  
see the features in figures \ref{fig4}b and c, within a separation of 3800~AU, 
as well as adjacent maser features within 100 AU. Figure \ref{fig3} shows 
that 10 out of 16 polarized features exhibited this tendency. The EVPAs were 
also constant at $-$13$\pm$5\deg\ in the individual features. The exception 
was found in the cluster of features seen in figure \ref{fig4}e, in which EVPAs 
changed by 90\deg. Only one feature at the south--west corner of the cluster, 
\w3:I2003 {\it 40}, had an EVPA roughly equal to those in other clusters. 

In W51M (LLD) and, EVPAs of masers are well aligned along the direction of 
a highly collimated flow. In \w3, on the other hand, EVPAs are not parallel to 
the direction of the whole outflows, which is oriented at $PA\: \sim$10\deg\ 
(see figure \ref{fig1} and figure 6 of Paper {\rm I}). In addition, they are not 
aligned in the direction of the large-scale 
CO outflow ($PA\: \sim$ 50\deg, \citealt{mit91,has94}). Note that the 
alignment of EVPAs observed in the present paper appears more evidently 
than the alignment of proper motions observed in Paper {\rm I}. This is 
probably because the random motions are dominant in the proper motions. 

\placefigure{fig5}

\section{Discussion}

\subsection{Relation between the EVPA and the magnetic field}
\label{sec:EVPA}

The present work has investigated for the first time the magnetic field of 
the whole water maser regions covering a large mapping area 
(2\arcsec$\times$5\arcsec) and the wide maser velocity range 
($\Delta V\:=$~53\kms). Despite the proper motions that have quite 
different PAs among the maser features, we found the similar EVPA values 
($-$13\deg.7($\pm$3\deg.7 in median) among 10 maser features widely 
distributed in the whole maser region. This position angle should be 
perpendicular or parallel to the magnetic field in the water maser region 
of \w3. 

A maser theory suggests that the directions of the linear polarization of 
saturated masers should be perpendicular (or parallel) to the magnetic 
field projection on the sky if the inclination angle of the magnetic field 
with respect to the line of sight is 
$|\theta_{M}|\:\gtrsim$~55\deg\ (or $|\theta_{M}|\:\lesssim$~55\deg) 
\citep{gol73,deg86b}. This inclination angle was roughly estimated from 
the observable parameter $Q/I$ (see figure 2 of \citealt{deg86a}). We 
estimated the inclination angle to be 
45\deg\ $\lesssim\; \theta_{\mbox{\tiny M}}\; \lesssim$ 60\deg\ for the 
observed value $|Q/I|\:\lesssim$~0.15, which is close to the critical 
inclination angle mentioned above ($\sim$~55\deg). We expect that 
the feature at the south--west corner of the cluster, \w3:I2003 {\it 40}, 
exhibits a "flip" of the linear polarization direction by $\sim$~90\deg\ 
because the magnetic field has an inclination angle close to the 
critical value. However, in other features, we favor 
$|\theta_{M}|\:\gtrsim$~55\deg\ because the magnetic field on the larger 
scale is almost tangential to the sky plane as mentioned above. Thus, the 
magnetic field projection was estimated to have a position angle 
$PA\:\sim$~76\deg. In contrast, the directions of linear polarization in 
the W51~M water maser are considered to be parallel to the magnetic field 
(LLD). 

Note that the linear polarization is affected more or less by Faraday rotation 
due to free electron in the interstellar medium along the line of sight. We 
evaluated Faraday rotation for the maser linear polarization at 1.35-cm wave 
length, $\Phi$, using the following qeuation,

\begin{equation}
\label{eq:Faraday}
\Phi\: \;\simeq\; 1.48\times 10^{2}Dn_{\mbox{\tiny e}}H_{\parallel} , 
\end{equation}

\noindent 
where $D$ (pc) is the size of an interstellar cloud, 
$n_{\mbox{\tiny e}}$ (cm$^{-3}$) the number density of free electron, 
and $H_{\parallel}$ (G) the magnetic field strength along the line of sight. 
For the cases of a maser cloud 
($D\:\sim$~10$^{-4}$~pc, $n_{\mbox{\tiny e}}\:\sim$~1~cm$^{-3}$, 
$H_{\parallel}\:\sim$~30~mG), a molecular cloud 
($D\:\sim$~0.1~pc, $n_{\mbox{\tiny e}}\:\sim$~1~cm$^{-3}$, 
$H_{\parallel}\:\sim$~100~$\mu$G), and the interstellar space 
between the W3 region and the Sun 
($D\:\sim$~1.8~kpc, $n_{\mbox{\tiny e}}\:\sim$~10$^{-2}$~cm$^{-3}$, 
$H_{\parallel}\:\sim$~10~$\mu$G), we obtain 
$\Phi\:\sim$~4.4$\times$10$^{-4}$~rad, 1.5$\times$10$^{-3}$~rad, 
and 2.7$\times$10$^{-2}$~rad, respectively, which are negligible effects. 
Although the hyper-compact H{\rm II} regions found in the \w3 region 
($D\:\sim$~2$\times$10$^{-3}$~pc, 
$n_{\mbox{\tiny e}}\:\sim$~1.3$\times$10$^{5}$~cm$^{-3}$, 
$H_{\parallel}\:\sim$~30~mG, \citealt{cla94}) might introduce the 
strongest Faraday rotation ($\Phi\:\sim$~120~rad), they are not located 
in front or behind the water maser features with linear polarization. 

\subsection{The magnetic field on the arcsecond-scale in \w3}
\label{sec:hourglass}

\citet{tro89}, RCTG, and \citet{gre94} found a gradient of the magnetic 
field along the line of sight and proposed "hourglass" models for the 
\w3 magnetic field on a large scale (1\arcmin). In these models, an 
inclination angle of the hourglass major axis $\theta\:\simeq$ 2\deg\ was 
adopted with respect to the sky plane on the basis of the angular 
distribution of the magnetic field strength along the line of sight, which 
was obtained by RCTG. A position angle of the major axis, 
$PA\: \sim$~50\deg, and a FWHM width of the pinch, 
$FWHM\:=$~10\arcsec, were adopted on the basis of the spatial 
distribution of the EVPAs, which was obtained by \citet{gre94} with 
an angular resolution of 13\arcsec.5. 

\begin{figure}
\epsscale{0.8}
\plotone{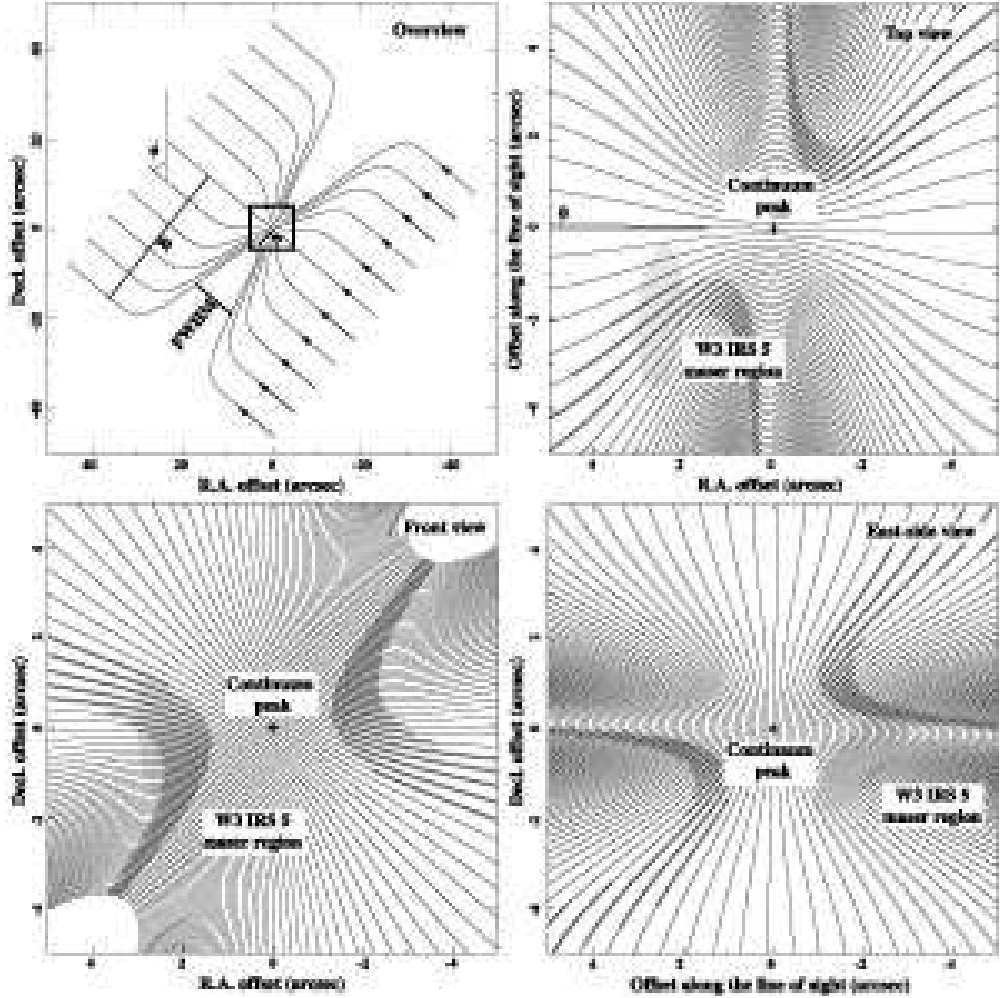}

\caption{Hourglass model of the magnetic field in the \w3 region. The angle 
$\theta\sim$~2\deg\ is the angle of inclination with respect to the sky plane, 
$\phi\sim$~50\deg\ the position angle, $A/B$ the pinch ratio 
($A\:=$~5\arcsec, $B\:=$~35\arcsec) and $FWHM\:=$~10\arcsec\ the FWHM 
width of the the pinch at the location of the water masers. The all of the 
definitions are the same as those described in figure 7 of Robert\etal\ 
(1993). A square (10\arcsec$\times$10\arcsec) in the overview panel 
indicates the area zoomed up and described in other three sub-panels. 
Thick parts of the magnetic field lines in the zoomed-up front view sub-panel 
indicate the locations where the field line have position angles 
65\deg\ $\leq\; PA_{\mbox{\tiny M}}\;\leq$~85\deg. The location and the 
area of the \w3 water maser region are indicated by a gray ellipse. 
The area center is located at 2\arcsec.5, $-$0\arcsec.5, and $-$2\arcsec.0 
in R.A., decl., and line-of-sight offsets, respectively. The orientation of 
the ellipse major axis indicates the direction of the outflow found in 
the maser region (see figure 6 and 7 of Paper {\rm I}). Thick parts of 
the magnetic field lines in the zoomed-up top and east-side views indicate 
the locations where the field line have inclinations 
45\deg~$\leq\; |\theta_{\mbox{\tiny M}}|\;\leq$~60\deg. 
\label{fig5}}
\end{figure}

In this paper, we propose an hourglass model of the magnetic field shown 
in figure \ref{fig5}. This model is made to explain the magnetic field on the 
small scale ($<$~10\arcsec), and consistent the large scale magnetic field 
($\sim$~1\arcmin) observed by RCTG and \citet{gre94}. The modeled field 
is pinched-in at the peak of the 800-\micron continuum emission. 
\citet{lad93} observed the 450-\micron/800-\micron  continuum 
emission with a 7\arcsec-beam (a pointing accuracy of 2\arcsec), and 
found that the peak of the emission has a position offset of 
$\sim$~$-$2\arcsec.5 in R.A.\ and $\sim$~$+$0\arcsec.5 in decl.\ with 
respect to the center of the maser cluster. Although the continuum 
emission peak should coincide with the $^{13}$CO and C$^{18}$O emission 
peaks \citep{old94, tie95, rob97}, the observed position is somewhat 
uncertain because of the large beams. It is possible that the C$^{13}$O 
emission peak observed by \citet{rob97} traces a segment of the outflow 
while the continuum emission peak traces the densest part of the parent 
molecular cloud. \citet{tie95} found four gas clumps traced by 
C$^{34}$S $J=$3--2 and $J=$5--4 emission, but none of the emission peaks 
coincides with the pinch center. 

The magnetic field projection on the sky ($PA\:\sim$~76\deg) mentioned 
in the previous sub-section is well consistent with that in the proposed 
magnetic field model (see figure \ref{fig5}b). Taking into account that the 
magnetic field vector is oriented toward the observer (a negative value of 
the measured Zeeman effect, STR; STCR), the maser region is likely 
located in front of the major axis of the hourglass (see figure \ref{fig5}c,~d). 
The inclination of the magnetic field to the line of sight, however, cannot 
be completely consistent with the model, in which the inclination 
significantly varies from a point to another in the whole maser region. 

The alignment of the magnetic field in the coverage of about 
1\arcsec$\times$4\arcsec\ in the water maser region provides a 
constraint on the pinch ratio of the hourglass model around 
$A/B\: \sim$~0.14, as well as on the pinch width around 
$FWHM\: \sim$~10\arcsec) (definitions of $A$ and $B$ are given in 
the caption of figure \ref{fig5}). As discussed by RCTG, one obtains a 
magnetic field strength of $\sim$1.3~mG at the pinch center when 
assuming a strength of the total undisturbed magnetic field of 
25~$\mu$G and considering the $r^{-2}$ dependence of the magnetic 
field. This value is one order of magnitude smaller than that obtained 
by STR and STCR ($\sim$~30~mG) but one order of magnitude larger 
than that estimated by RCTG. \citet{cru99} found an empirical relation 
of a magnetic field strength, $B\:\propto\: n^{0.47}$, over a 
large range of gas densities $n$ up to 10$^{7}$~cm$^{-3}$. Therefore, 
it is expected that the field strength found in the water maser cloud 
($n\:\gtrsim$~10$^{8}$--10$^{9}$~cm$^{-3}$) is consistent with the 
$B$--$n$ law (c.f.\ STR). STCR also pointed out that the magnetic field 
in the interstellar space (preshock region) should be enhanced in the 
maser region (postshock region) by a factor of $\sim$~20. 

\subsection{The magnetodynamically super-critical collapse of the 
parent molecular cloud of \w3}
\label{sec:collapse}

The origin of the magnetic field pinch is still obscured. The direction of the 
whole magnetic field is almost parallel to a molecular outflow found in W3~A 
\citep{mit91, has94}. \citet{mom01} pointed out that a magnetic field would 
control the direction of a molecular outflow rather than that the magnetic 
field is influenced by the outflow. In \w3, however, although the magnetic 
field seems to along the molecular outflow on a larger scale 
($\sim$~1\arcmin), it may not necessarily be parallel to the outflows on 
a smaller scale ($\sim$~5\arcsec) as seen in the water masers. We evaluated 
the energy densities of the magnetic field and of the outflow of \w3 using 
the same equations described in \citet{mom01}. The critical magnetic field 
strength, $B_{\mbox{\tiny critical}}$, at which the field has the same 
energy density as that of the cylindrical outflow, is expressed by  

\begin{equation}
\label{eq:bflow}
B_{\mbox{\tiny critical}}\;=\left(\frac{8E_{\mbox{\tiny flow}}}{r^{2}l}\right)^{1/2}, 
\end{equation}

\noindent 
where $E_{\mbox{\tiny flow}}$, $r$ and $l$ are the kinetic energy, the 
radius, and the total length of the cylindrical outflow, respectively. 
Adopting the total outflow energy estimated by \citet{has94}, 
$E_{\mbox{\tiny flow}}\:\simeq$~1.2$\times$10$^{46}$~erg, and the radius 
and the length of the outflow estimated by \citet{mit91}, 
$r\:\simeq$~4.1$\times$10$^{17}$~cm 
and $l\:\simeq$~2.5$\times$10$^{18}$~cm, respectively (assuming the 
distance to \w3 of 1.8~kpc here), we obtained 
$B_{\mbox{\tiny critical}}\:\simeq$~0.48~mG, 
which is larger than the maximum strength observed in the H{\rm I} Zeeman 
effect by RCTG ($\lesssim$~0.22~mG). To obtain $B_{\mbox{\tiny critical}}$ for 
the water maser cloud, we converted equation (\ref{eq:bflow}) to, 

\begin{equation}
\label{eq:bflow2}
B_{\mbox{\tiny critical}}\;=\; 
(8\pi\rho_{\mbox{\tiny maser}}v_{\mbox{\tiny maser}}^{2})^{1/2}, 
\end{equation}

\noindent
where $\rho_{\mbox{\tiny maser}}\:\gtrsim$
~3.3$\times$10$^{-14}$~g~cm$^{-3}$ (corresponding to 
$n_{\mbox{\tiny H$_{2}$}}\: =$~10$^{8}$~cm$^{-3}$) is the gas density of the 
maser region necessary for maser excitation, $v_{\mbox{\tiny maser}}$ is the 
outflow velocity. Adopting the outflow velocity 
$v_{\mbox{\tiny maser}}\:\sim$~15\kms, we obtained 
$B_{\mbox{\tiny critical}}\:\simeq$~250~mG. Thus, the magnetic field 
strength estimated on the basis of the Zeeman splitting ($\simeq$~30~mG, 
STR) is one order of magnitude lower than the critical strength. This 
suggests that the magnetic cannot control the outflow dynamics. 

This evaluation is consistent with that made for the W51M region by LLD. 
On the other hand, there is no observational evidence that the outflow 
controls the magnetic field structure. \citet{mom01} suggested that the 
magnetic field is aligned more in the younger phase of star-formation 
than in the more evolved one so that the disturbance of the former 
magnetic field due to the small-scale turbulence is not significant yet. 
This suggestion supports that \w3 should be at the very early stage of 
star formation. \citet{old94} and \citet{tie95} estimated the mass of the 
molecular cloud associated with \w3 to be 800--900~$M_{\sun}$. This 
value is still larger than the magnetodynamically-critical mass of 
510~$M_{\sun}$ for a cloud with a radius of 9000~AU (corresponding 
to 5\arcsec\ at 1.8~kpc) and a magnetic field strength of 30~mG. This 
indicates that the magnetic field of the \w3 cannot support the gravity 
of the molecular cloud, in other words, that a part of the \w3 cloud is 
collapsing under the magnetodynamically super-critical condition 
\citep{shu87}. 

Note that the above evaluation is still consistent with the suggestion that 
the magnetic field should control the water maser excitation (\citealt{eli89}, 
1992; STCR). To consider the maser excitation, the velocity dispersion 
of the postshock region where the masers are excited, 
$\sigma_v\:\lesssim$~0.6\kms, is inserted instead of 
$v_{\mbox{\tiny maser}}$ in equation \ref{eq:bflow2} 
(Paper {\rm I, II}; STCR). Then one can estimate 
$B_{\mbox{\tiny maser}}\:\simeq$~10~mG, smaller than or equal to the 
observed magnetic field strength. 

It is still unclear what kind of object exists in the pinch center of the 
magnetic field. \citet{tie95} found four molecular gas clumps in \w3, but 
none of the clumps is located at the pinch center. It is likely that these 
clumps were formed in the common parent molecular cloud and are 
responsible for the observed magnetic field. The magnetic field which 
controlled the multiple clumps supports the mild pinch ratio 
$A/B\:\simeq$~0.14 rather than the extremely small ratio 
$A/B\: \simeq$~0.04 speculated by RCTG. 

\section{Summary}

Polarimetric VLBI observations of water masers in \w3 have revealed that 
observed EVPAs of water masers features are well aligned in the whole 
maser region except for a small number of maser features. The directions 
of the magnetic field projected on the sky in the maser region are estimated 
to be perpendicular to the directions of the linear polarization of the 
masers ($PA\:\sim$~76\deg). The magnetic field estimated supports the 
hourglass magnetic field model proposed previously. Thus, water maser 
linear polarization may reveal in detail the magnetic fields of the interstellar 
medium in star-forming regions. The \w3 water maser region has a position 
offset from the pinch center of the hourglass and its magnetic field is 
unlikely to control the outflow dynamics. All of these observational facts 
suggest that the W3 IRS 5 cloud is collapsing under magnetically 
super-critical condition and that the outflows are not yet developed 
enough to completely disturb the magnetic field in this cloud. The 
hourglass structure of the magnetic field is suggested to be formed as a 
result of the collapse of the parent molecular cloud that is magnetically 
super-critical. 

\acknowledgments

Authors acknowledge to Dr.\ Chris Carilli for helping our VLBA and VLA 
observations as a technical contact person. Authors also thanks Drs.\ 
Tetsuo Sasao, Makoto Miyoshi, and Yoshiharu Asaki for helping the VLBA 
observation, and Dr.\ Motohide Tamura for providing fruitful information 
on interstellar polarimetric data. H.~I.\ was financially supported by the 
Research Fellowship of the Japan Society for the Promotion of Science for 
Young Scientist.

\clearpage

\input{tab.tex}

\clearpage

\end{document}

%% file: tab.tex
\begin{table}
\caption{Parameters of the water maser features toward \w3.
\label{tab1}}
{\tiny
\begin{tabular}{lrrrrrrrc} \hline \hline
Maser feature\footnotemark[1] 
& \multicolumn{2}{c}{Offset\footnotemark[2]} 
& \multicolumn{2}{c}{LSR velocity (km s$^{-1}$)} 
& \multicolumn{1}{c}{Peak Intensity}
& \multicolumn{1}{c}{$P$\footnotemark[4]} & \multicolumn{1}{c}{EVPA} 
& Maser feature\footnotemark[5] \\
 (\w3: I2003) & R.A. (mas) & decl. (mas) & Peak & Width\footnotemark[3] 
& (Jy beam$^{-1}$) 
& \multicolumn{1}{c}{(\%)} & (deg) & (\w3: I2000) \\ \hline
  1   \ \dotfill \  &    $-$1.50 &      3.33 & $-$56.0 &  1.7 
 &      0.04 &  ...  &  ... & ...  \\
  2   \ \dotfill \  &   1075.72 &   1016.29 & $-$43.1 &  3.8
 &     18.41 & 11.0 & $-$12.1 & ... \\
  3   \ \dotfill \  &  $-$389.93 &  $-$661.34 & $-$42.9 & 1.7
 &      7.78 &  ...  &  ... & ... \\
  4   \ \dotfill \  &      0.08 &    $-$0.99 & $-$42.6 &  1.3
 &      6.95 &  ...  &  ...  & ...  \\
  5   \ \dotfill \  &    $-$1.49 &      3.36 & $-$42.1 &  3.8
 &    111.05 & 0.8 &  $-$15.1 & ...  \\
  6   \ \dotfill \  &  $-$388.32 &  $-$665.56 & $-$42.0 &  2.1
 &      9.60 & 0.7 &   19.8 & ... \\
  7   \ \dotfill \  &  $-$388.08 &  $-$671.67 & $-$41.7 &  0.8
 &      3.63 &  ...  &  ... & ...   \\
  8   \ \dotfill \  &  $-$431.85 &  $-$620.46 & $-$41.7 &  0.8
 &      1.48 &  ...  &  ... & 29   \\
  9   \ \dotfill \  &  $-$387.87 &  $-$667.89 & $-$41.3 &  0.4
 &      0.50 &  ...  &  ...  & ...  \\
 10   \ \dotfill \  &    $-$0.01 &    $-$0.09 & $-$41.3 &  1.3
 &     26.10 &  ...  &  ... & ...   \\
 11   \ \dotfill \  &   $-$18.42 &      0.16 & $-$40.7 &  1.7
 &      8.95 &  ...  &  ... & ...   \\
 12   \ \dotfill \  &   $-$17.26 &    $-$1.79 & $-$40.5 &  1.3
 &      2.63 &  ...  &  ... & ...   \\
 13   \ \dotfill \  &    $-$1.37 &      4.38 & $-$40.5 &  1.7
 &      8.69 &  ...  &  ... & ...   \\
 14   \ \dotfill \  &      0.00 &      0.00 & $-$40.5 &  5.1
 &     79.89 & 0.8 & $-$14.9 & ...  \\
 15   \ \dotfill \  &    $-$1.66 &      5.47 & $-$40.3 &  1.3
 &      8.91 &  ...  &  ... & ...   \\
 16   \ \dotfill \  &    $-$0.37 &   $-$11.64 & $-$40.0 &  2.5
 &     20.81 &  ...  &  ... & ...   \\
 17   \ \dotfill \  &    $-$3.92 &      5.23 & $-$39.6 &  0.4
 &      5.05 &  ...  &  ... & ...   \\
 18   \ \dotfill \  &  $-$387.50 &  $-$670.38 & $-$39.3 &  4.2
 &      9.99 & 0.8 & $-$17.8 & ...  \\
 19   \ \dotfill \  &  $-$386.67 &  $-$677.37 & $-$38.9 &  3.0
 &      7.25 &  ...  &  ... & ...   \\
 20   \ \dotfill \  &    $-$4.28 &      8.23 & $-$38.8 &  1.3
 &      5.63 &  ...  &  ...  & ...  \\
 21   \ \dotfill \  &    $-$3.78 &      7.50 & $-$38.8 &  0.8
 &      3.91 &  ...  &  ... & ...   \\
 22   \ \dotfill \  &    $-$6.27 &      8.67 & $-$38.2 &  3.8
 &     26.49 & 0.7 & $-$11.8 & ...  \\
 23   \ \dotfill \  &      9.72 &   $-$16.39 & $-$37.8 &  2.1
 &     11.11 & 12.8 & $-$11.4 & ...  \\
 24   \ \dotfill \  &  $-$386.73 &  $-$676.74 & $-$37.7 &  1.3
 &      0.29 &  ...  &  ... & ...   \\
 25   \ \dotfill \  &    $-$7.97 &     13.24 & $-$37.5 &  1.7
 &      1.88 &  ...  &  ... & ...   \\
 26   \ \dotfill \  &    $-$9.20 &     14.11 & $-$36.9 &  1.7
 &      2.14 &  ...  &  ... & ...   \\
 27   \ \dotfill \  &    786.55 &   1792.69 & $-$36.7 &  2.1
 &     10.01 & 1.1 & $-$15.5 & 64  \\
 28   \ \dotfill \  &    797.51 &   1750.03 & $-$36.2 &  0.8
 &      0.56 &  ...  &  ... & 65   \\
 29   \ \dotfill \  &    783.13 &   1802.02 & $-$35.8 &  3.8
 &     14.81 & 0.9 & $-$18.0 & 66  \\
 30   \ \dotfill \  &    787.03 &   1794.17 & $-$35.5 &  5.1
 &     91.10 & 0.4 & $-$12.5 & 63  \\
 31   \ \dotfill \  &    328.54 & $-$2001.74 & $-$35.0 &  1.7
 &      2.24 &  ...  &  ... & 71   \\
 32   \ \dotfill \  &    796.47 &   1750.38 & $-$35.0 &  3.0
 &      0.53 &  ...  &  ... & ...  \\
 33   \ \dotfill \  &  $-$416.12 &  $-$600.11 & $-$34.6 & 4.2
 &     11.55 &  ...  &  ... & ...  \\
 34   \ \dotfill \  &    $-$1.46 &      3.30 & $-$34.6 &  0.4
 &      0.11 &  ...  &  ... & ...  \\
 35   \ \dotfill \  &    $-$0.02 &    $-$0.02 & $-$34.6 & 0.4
 &      0.06 &  ...  &  ... & ...  \\
 36   \ \dotfill \  &    $-$6.21 &      8.51 & $-$34.6 &  0.4
 &      0.08 &  ...  &  ... & ...  \\
 37   \ \dotfill \  &    794.01 &   1751.23 & $-$34.5 &  0.4
 &      2.10 &  ...  &  ... & ...  \\
 38   \ \dotfill \  &    794.65 &   1750.92 & $-$34.1 &  0.4
 &      0.23 &  ...  &  ... & 79  \\
 39   \ \dotfill \  &    295.92 & $-$2025.00 & $-$33.9 &  1.7
 &      0.94 &  ...  &  ... & 73  \\
 40   \ \dotfill \  &    294.72 & $-$2025.82 & $-$33.8 &  2.5
 &      4.43 & 5.4 & $-$5.0 & 76 \\
 41   \ \dotfill \  &    293.69 & $-$2027.11 & $-$33.7 &  0.8
 &      1.06 &  ...  &  ... & ...  \\
 42   \ \dotfill \  &    331.95 & $-$2004.59 & $-$33.7 &  1.3
 &      6.95 & 7.0 & 89.6 & ... \\
 43   \ \dotfill \  &    325.76 & $-$1999.60 & $-$33.6 &  4.6
 &     21.89 & 7.3 & 30.4 & 82 \\
 44   \ \dotfill \  &    332.52 & $-$2004.95 & $-$33.3 &  0.8
 &      4.25 & 12.9 & $-$79.2 & ... \\
 45   \ \dotfill \  &  $-$415.21 &  $-$602.42 & $-$33.2 &  3.0
 &      1.00 &  ...  &  ... & ...  \\
 46   \ \dotfill \  &    334.25 & $-$2006.04 & $-$33.1 &  3.0
 &     10.90 & 8.4 & $-$59.1 & 83  \\
 47   \ \dotfill \  &  $-$415.45 &  $-$604.25 & $-$33.0 &  3.4
 &      3.84 &  ...  &  ... & ...  \\
 48   \ \dotfill \  &    324.56 & $-$1998.90 & $-$32.9 &  5.9
 &     13.12 &  ...  &  ... & ...  \\
 49   \ \dotfill \  &    324.99 & $-$1998.91 & $-$32.4 &  2.1
 &      6.56 & 1.5 & $-$62.5 & ... \\
 50   \ \dotfill \  &    795.87 &   1750.48 & $-$32.3 &  2.1
 &      1.93 &  ...  &  ... & ...  \\
 51   \ \dotfill \  &    326.65 & $-$1999.60 & $-$32.0 &  1.3
 &      0.19 &  ...  &  ... & 77  \\
 52   \ \dotfill \  &    331.83 & $-$2004.52 & $-$31.2 &  0.4
 &      0.05 &  ...  &  ... & ...  \\
 53   \ \dotfill \  &    249.87 & $-$2079.93 & $-$30.4 &  4.6
 &      4.02 &  ...  &  ... & 91  \\
 54   \ \dotfill \  &    228.02 & $-$2083.62 & $-$29.9 &  1.7
 &      1.05 &  ...  &  ... & 94  \\
 55   \ \dotfill \  &    230.35 & $-$2083.38 & $-$29.9 &  0.8
 &      0.44 &  ...  &  ... & 98  \\
 56   \ \dotfill \  &    229.26 & $-$2083.48 & $-$29.9 &  0.4
 &      0.01 &  ...  &  ... & ... \\
 57   \ \dotfill \  &    227.05 & $-$2083.52 & $-$29.9 &  1.7
 &      1.09 &  ...  &  ... & ... \\
 58   \ \dotfill \  &    325.72 & $-$1999.57 & $-$29.6 &  2.1
 &      0.06 &  ...  &  ... & ... \\
 59   \ \dotfill \  &    795.39 &   1750.80 & $-$29.5 &  1.3
 &      0.13 &  ...  &  ... & ... \\
 60   \ \dotfill \  &  $-$503.83 & $-$2377.04 & $-$23.8 & 3.0
 &      0.58 &  ...  &  ... & ... \\
 61   \ \dotfill \  &  $-$205.57 & $-$1830.29 &  $-$6.8 & 2.1
 &      1.32 &  ...  &  ... & ...  \\
 \hline
 \end{tabular}
}

\tiny \noindent
\footnotemark[1] Water maser features detected toward \w3. 
The feature is designated as \w3:I2003 {\it N}, where {\it N} is 
the ordinal source number given in this column (I2003 
stands for sources found by Imai\etal\ and listed in 2003). \\
\footnotemark[2] Relative value with respect to the location of the 
position-reference maser feature: \w3:I2003 {\it 14}. \\
\footnotemark[3] A channel velocity spacing multiplied by a number 
of detected maser spots in the feature. \\
\footnotemark[4] Fraction of linear polarization intensity to total intensity. \\
\footnotemark[5] The feature identical to that found in Paper {\rm I}, \w3:I2000 
{\it M}, where {\it M} is the ordinal source number given in this column. \\

\end{table}

%% file: ms.bbl
\clearpage
\begin{thebibliography}{}
%\setlength{\baselineskip}{1.2zh}
\bibitem[Abraham \&  Vilas Boas(1994)]{abr94}
Abraham,~Z., \&  Vilas Boas,~J.~W.~S.\  1994, \aap, 290, 956 
\bibitem[Akujor\etal(1993)]{aku93}
Akujor,~C.~E., Spencer,~R.~E., Zhang,~F.~J., Fanti,~C., Ludke,~E., \&  
Garrington,~S.~T.\  1993, \aap, 274, 752
\bibitem[Barvainis \&  Deguchi(1989)]{bar89}
Barvainis,~R., \&  Deguchi,~S.\ 1989, \aj, 97, 1089
\bibitem[Crutcher(1999)]{cru99}
Crutcher,~R.~M.\ 1999, \apj, 520, 706
\bibitem[Claussen\etal(1994)]{cla94}
Claussen,~M.~J., Gaume,~R.~A., Johnston,~K.~J., 
\&  Wilson,~T.~L.\  1994, \apjl, 424, L41
\bibitem[Deguchi \&  Watson(1990)]{deg90}
Deguchi,~S., \&  Watson,~W.~D.\ 1990, \apj, 354, 649
\bibitem[Deguchi \&  Watson(1986a)]{deg86a}
--.\ 1986a, \apj, 302, 750
\bibitem[Deguchi \&  Watson(1986b)]{deg86b}
--.\ 1986b, \apj, 302, 108
\bibitem[Diamond(1995)]{dia95}
Diamond,~P.~J.\  1995, in ASP Conf.~Ser.~82, VERY LONG BASELINE 
INTERFEROMETRY AND THE VLBA, ed.\ J.~A.~Zensus, P.~J.~Diamond, 
\&  P.~J.~Napier (San Francisco: ASP), 227
\bibitem[Elitzur(1992)]{eli92a} 
Elitzur,~M.\ 1992a, in Astronomical Masers (Dordrecht: Kluwer)
\bibitem[Elitzur(1992)]{eli92b} --.\ 1992b, \araa, 30, 75
\bibitem[Elitzur, Hollenbach, \&  McKee(1989)]{eli89} 
Elitzur,~M., Hollenbach,~D.~J., \&  McKee,~C.~F.\  1989, \apj, 346, 983
\bibitem[Elitzur, Hollenbach, \&  McKee(1992)]{eli92c}
--.\ 1992, \apj, 394, 221
\bibitem[Evans~II(1999)]{eva99} Evans~II,~N.~J.\ 1999, \araa, 37, 311
\bibitem[Feigelson \&  Montmerle(1999)]{fei99}
Feigelson,~E.~D., \&  Montmerle,~T.\  1999, \araa, 37, 363
\bibitem[Goldreich, Keeley, \& Kwan(1973)]{gol73}
Goldreich,~P., Keeley,~D.~A., \& Kwan,~J.~Y.\ 1973, \apj, 179, 111
\bibitem[Greaves, Murray, \&  Holland(1994)]{gre94} 
Greaves,~J.~S., Murray,~A.~G., \&  Holland,~W.~S.\  1994, \aap, 284, L19
\bibitem[Genzel \&  Downes(1977)]{gen77} 
Genzel,~R., \&  Downes,~D.\  1977, \aaps, 30, 145
\bibitem[Gwinn(1994)]{gwi94a} Gwinn,~C.~R.\  1994a, \apj, 429, 241
\bibitem[Gwinn(1994)]{gwi94b} --.\  1994b, \apj, 429, 253
\bibitem[Gwinn, Moran, \&  Reid(1992)]{gwi92} 
Gwinn,~C.~R., Moran,~J.~M., \&  Reid,~M.~J.\  1992, \apj, 393, 149
\bibitem[Hasegawa\etal(1994)]{has94}
Hasegawa,~T.~I., Mitchell,~G.~F., Mattews,~H.~E., \&  
Tacconi,~L.\  1994, \apj, 426, 215
\bibitem[Hayashi, Kobayashi, \& Hasegawa(1989)]{hay89}
Hayashi,~M., Kobayashi,~H., \&  Hasegawa,~.T.\ 1989, \apj, 340 298
\bibitem[Heiles\etal(1993)]{hei93}
Heiles,~C., Goodman,~A.~A., McKee,~C.~F., \&  Zweibel,~E.G.\ 1993, 
in Protostar and Planets III, ed.\ E.~H.~Levy,  \&  J.~I.~Lunine 
(Tucson: Univ.\ Arizona Press.), 279
\bibitem[Hollenbach, Elitzur, \&  McKee(1993)]{hol93}
Hollenbach,~D.~J., Elitzur,~M., \&  McKee,~C.~F.\  1993, 
in Astrophysical Masers, ed.\ A.~W.~Clegg, \& G.~E.~Nedoluha 
(Heidelberg: Springer), 159
\bibitem[Hollenbach \& McKee(1989)]{hol89}
Hollenbach,~D., \&  McKee,~C.~F.\  1989, \apj, 342, 306
\bibitem[Horiuchi, Migenes, \&  Deguchi(2000)]{hor00a}
Horiuchi,~S., Migenes,~V., \&  Deguchi,~S.\ 2000, in Astrophysical Phenomena 
Revealed by Space VLBI, Proceedings of the VSOP Symposium, eds.\ 
H.~Hirabayashi, P.~G.~Edwards, \&  D.~W.~Murphy, (Sagamihara, Japan: ISAS), 105
\bibitem[Horiuchi \&  Kameya(2000)]{hor00b}
Horiuchi,~S., \& Kameya,~O.\  2000, \pasj, 52, 545
\bibitem[Imai, Deguchi, \&  Sasao(2002)]{ima02}
Imai,~H., Deguchi,~S., \&  Sasao,~T.\ 2002, \apj, 567, 971 (Paper {\rm II})
\bibitem[Imai \etal(2000)]{ima00}
Imai,~H., Kameya,~O., Sasao,~T., Miyoshi,~M., Deguchi,~S., Horiuchi,~S., 
\&  Asaki,~Y.\  2000, \apj, 538, 751 (Paper {\rm I})
\bibitem[Ladd\etal(1993)]{lad93}
Ladd,~E.~F., Deane,~J.~R., Sanders,~D.~B., \& Wynn-Williams,~C.~G.\ 1993, 
\apj, 419, 186
\bibitem[Lepp\"anen, Liljestr\"om, \&  Diamond(1998)]{lep98}
Lepp\"anen,~K., Liljestr\"om,~T., \&  Diamond,~P.J.\  1998, 
\apj, 507, 909 (LLD)
\bibitem[Marscher\etal(2002)]{mar02}
Marscher,~A.~P., Jorstad,~S.~G., Mattox,~J.~R., \&  Wehrle,~A.\ 2002, 
\apj, 577, 85
\bibitem[McKee\etal(1993)]{mck93}
McKee,~C.~F., Zweobe,~E.~G., Godman,~A.~A., \&  Heiles,~C.\ 1993, 
in Protostar and Planets III, ed.\ E.~H.~Levy,  \&  J.~I.~Lunine 
(Tucson: Univ.\ Arizona Press.), 327
\bibitem[Mitchell, Maillard, \&  Hasegawa(1991)]{mit91}
Mitchell,~G.~F., Maillard,~J.-P., \&  Hasegawa,~T.~I.\ 1991, \apj, 371, 342
\bibitem[Momose\etal(2001)]{mom01}
Momose,~M., Tamura,~M., Kameya,~O., Greaves,~J.~S., Chrysotomou,~A., 
Hough,~J.~H., \&  Morino,~J.-I.\  2001, \apj, 555, 855
\bibitem[Oldham\etal(1994)]{old94}
Oldham,~P.~G., Griffin,~M.~J., Richardson,~K.~J., \&  Sandell,~G.\ 1994, 
\aap, 284, 559
\bibitem[Reid(1995)]{rei95} 
Reid,~M.~J.\  1995, in ASP Conf.~Ser.~Vol.~82, Very Long Baseline 
Interferometry and the VLBA, ed.\ J.~A.~Zensus, P.~J.~Diamond, \&  
P.~J.~Napier (San Francisco: ASP), 209
\bibitem[Reid \&  Moran(1981)]{rei81} 
Reid,~M.~J., \&  Moran,~J.~M.\  1981, \araa, 19, 231
\bibitem[Roberts, Crutcher, \&  Troland(1997)]{rob97}
Roberts,~D.~A., Crutcher,~R.~M., \&  Troland,~T.~H.\  1997, \apj, 479, 318
\bibitem[Roberts\etal(1993)]{rob93}
Roberts,~D.~A., Crutcher,~R.~M., Troland,~T.~H., \&  
Goss,~W.~M.\  1993, \apj, 412, 675 (RCTG)
\bibitem[Sarma\etal(2002)]{sar02}
Sarma, A.~P., Troland, T.~H., \&  Crutcher,~R.~M., \&  Roberts,~D.~A.\  
2002, \apj, 580, 928 (STCR)
\bibitem[Sarma, Troland, \& Romney(2001)]{sar01}
Sarma, A.~P., Troland, T.~H., \&  Romney, J.~D.\  2001, \apj, 554, 
L217 (STR)
\bibitem[Shu, Adams, \& Lizano(1987)]{shu87}
Shu,~F.~H., Adams,~F.~C., \&  Lizano,~S.\  1987, \araa, 25, 23
\bibitem[Tieftrunk\etal(1997)]{tie97}
Tieftrunk, A.~R., Gaume, R.~A., Claussen, M.~J., Wilson, T.~L., \&  
Johnston, K.~J.\  1997, \aap, 318, 931
\bibitem[Tieftrunk\etal(1995)]{tie95}
Tieftrunk, A.~R., Wilson, T.~L., Steppe,~H, Gaume, R.~A., Johnston, K.~J., 
\& Claussen, M.~J.\  1995, \aap, 303, 901
\bibitem[Torrelles\etal(2001)]{tor01a}
Torrelles,~J.~M., Patel,~N.~A., G\'omez,~J.~F., Ho,~P.~T.~P., 
Rodor\'iguez,~L.~F., Anglada,~G., Garay,~G., Greenhill,~L.~J., \etal\  
2001a, \nat, 411, 277
\bibitem[Torrelles\etal (2001)]{tor01b}
--.\ 2001b, \apj, 560, 853
\bibitem[Troland\etal(1989)]{tro89}
Troland,~T.~H., Crutcher,~R.~M., Goss,~W.~M., \&  Heiles,~C.\ 1989, 
\apj, 347, L89
\end{thebibliography}
